\def\la{\hbox{{\lower -2.5pt\hbox{$<$}}\hskip -8pt\raise
-2.5pt\hbox{$\sim$}}}
\def\ga{\hbox{{\lower -2.5pt\hbox{$>$}}\hskip -8pt\raise
-2.5pt\hbox{$\sim$}}}
\def\ltsima{$\; \buildrel < \over \sim \;$}
\def\simlt{\lower.5ex\hbox{\ltsima}}
\def\gtsima{$\; \buildrel > \over \sim \;$}
\def\simgt{\lower.5ex\hbox{\gtsima}}

\documentstyle[epsfig]{elsart}

\begin{document}
\begin{frontmatter}
\title{A fluctuating energy-momentum may produce an unstable world}
\author[lngs]{R. Aloisio},
\author[inaf]{P. Blasi},
\author[dipaq]{A. Galante},
\author[lngs]{A.F. Grillo}
\address[lngs]{INFN - Laboratori Nazionali del Gran Sasso, SS. 17bis\\
67010 Assergi (L'Aquila) - Italy}
\address[inaf]{INAF - Osservatorio Astrofisico di Arcetri, Largo E. Fermi 5\\
50125 Firenze - Italy}
\address[dipaq]{Dipartimento di Fisica, Universit\`a di L'Aquila, Via Vetoio\\
67100 Coppito (L'Aquila) - Italy}

\begin{abstract}
Quantum gravitational effects may induce stochastic fluctuations in the
structure of space-time, to produce a characteristic foamy structure. 
It has been known for some time now that these fluctuations may have
observable consequencies for the propagation of cosmic ray particles
over cosmological distances. We note here that the same fluctuations,
if they exist, imply that some decay reactions normally forbidden by 
elementary conservation laws, become kinematically allowed, inducing 
the decay of particles that are seen to be stable in our universe. Due 
to the strength of the prediction, we are led to consider this finding 
as the most severe constraint on the classes of models that may describe 
the effects of gravity on the structure of space-time.
 We also propose and discuss several potential loopholes of our approach, 
that may affect our conclusions. In particular, we try to identify the
situations in which despite a fluctuating energy-momentum of the
particles, the reactions mentioned above may not take place.
\end{abstract}
\end{frontmatter}

\section{Introduction}

In the last few years the hunt for possible minuscule violations of the
fundamental Lorentz invariance (LI) has been object of renewed interest,
in particular because it has been understood that cosmic ray physics has
an unprecedented potential for investigation in this field
\cite{kir,lgm,cam,colgla,noi,spain}. Some authors \cite{cam,colgla,berto}
have even invoked possible violations of LI as a plausible explanation to
some puzzling observations related to the detection of ultra high energy
cosmic rays (UHECRs) with energy above the so-called GZK feature \cite{gzk}, 
and to the unexpected shape of the spectrum of photons with super-TeV 
energy from sources at cosmological distances. 
Both types of observations have in fact 
many uncertainties, either coming from limited statistics of very rare events,
or from accuracy issues in the energy determination of the detected 
particles, and most likely the solution to the alleged puzzles will come from 
more accurate observations rather than by a violation of fundamental 
symmetries. 
For this reason, from the very beginning we proposed \cite{noi} that 
cosmic ray observations should be used as an ideal tool to constrain 
the minuscule violations of LI, rather than as evidence for the need 
to violate LI. The reason why the cases of UHECRs and TeV gamma rays 
represent such good test sites for LI is that both are related to physical
processes with a kinematical energy threshold, which is in turn very sensitive
to the smallest violations of LI. UHECRs are expected to suffer severe 
energy losses due to photopion production off the photons of the cosmic 
microwave background (CMB), and this should suppress the flux of particles 
at the Earth at energies above $\sim 10^{20}$ eV, the so called GZK feature. 
Present operating experiments are AGASA
and HiRes, and they do not provide strong evidence either in favor or 
against the detection of the GZK feature \cite{demarco}. A substantial 
increase in the statistics of events, as expected with the Auger project
and with EUSO should dramatically change the situation and allow to detect
the presence or lack of the GZK feature in the spectrum of UHECRs. These
are the observations that will provide the right ground for imposing a 
strong limit on violations of LI. For the case of TeV sources, the process
involved is pair production \cite{gamgam} of high energy gamma rays on
the photons of the infrared background. In both cases, a small violation
of LI can move the threshold to energies which are smaller than the 
classical ones, or move them to infinity, making the reactions impossible.
The detection of the GZK suppression or the cutoff in the gamma ray 
spectra of gamma ray sources at cosmological distances will prove that 
LI is preserved to correspondingly high accuracy \cite{noi}.

The recipes for the violations of LI generally consist of requiring an 
{\it explicit} modification of the dispersion relation of high energy 
particles, due to their propagation in the ``vacuum'', now affected by 
quantum gravity (QG). This effect is generally parametrized by introducing 
a typical mass, expected to be of the order of the Plank mass ($M_P$), 
that sets the scale for QG to become effective. 

However, explicit modifications of the dispersion relation are not really
necessary in order to produce detectable effects, as was recently
pointed out in Refs. \cite{ford,ng1,ng2,lieu} for the case of propagation 
of UHECRs. It is in fact generally believed that coordinate measurements 
scannot be performed with precision better than the Planck distance (time) 
$\delta x \geq l_P$, namely the distance where the metric of space-time 
must feature quantum fluctuations.
A similar line of thought implies that an uncertainty in the measurement
of energy and momentum of particles can be expected, according with the  
relation $\delta p \simeq \delta E \simeq p^2/M_P$ 

The consequences of these uncertainty relations on the propagation of 
high energy cosmic rays have been investigated by many authors. In a
previous paper \cite{noi2} (see also \cite{lastcamel}) we have actually 
argued that the effects of fluctuating dispersion relations may induce
observable consequences on cosmic ray propagation even at energies as
low as $\approx 10^{15}$ eV.

Very recently it has been argued (\cite{lieu2}, see also \cite{rag}) 
that generic fluctuations 
may already be inconsistent with detection of interference features 
from very distant sources; this result however has been criticized in 
\cite{ng3}.

In this paper we derive another important consequence of the fluctuating
dispersion relations introduced in \cite{noi2}: a particle propagating 
in vacuum acquires an energy dependent fluctuating effective mass, which
may be responsible for kinematically forbidden decay reactions
to become kinematically allowed. If this happens, particles that are known 
to be stable would decay, provided no other fundamental conservation law
is violated (e.g.: baryon number conservation, charge conservation).
A representative example is that of the reaction $p\to p+\pi^0$, that 
is prevented from taking place only due to energy conservation. 
With a fluctuating
metric, we find that if the initial proton has energy above a few $10^{15}$
eV, the reaction above can take place with a cross section typical of 
hadronic interactions, so that the proton would rapidly lose its energy. 
Similar conclusions hold for the electromagnetic process $p\to p+\gamma$.

The fact that particles that would be otherwise stable could
decay has been known for some time now {\cite{gmestres,liberati} and
in fact it rules out a class of non-fluctuating modifications of the 
dispersion relations for some choices of the sign of the modification: 
the new point here is that it does not appear to be possible to fix the 
sign of the fluctuations, so that the conclusions illustrated above 
seem unavoidable.

The plan of the paper is the following: in \S 1 we discuss our
main results, putting strong emphasys on the underlying assumptions.
 In \S 2, we set the framework for kinematical computations of 
the thresholds for the processes presented in \S 3. Finally in the last
section we argue that the comparison of our predictions with experimental 
data indicates a strong inconsistency, implying that the framework of 
quantum fluctuations currently discussed in most literature is in fact 
ruled out. The strength of this conclusion leads us to try to identify 
possible loopholes in our working assumptions. The ways to avoid the
dramatic effects of the fluctuating energy-momentum of a particle should
be mainly searched in the dynamics of Quantum Gravity.
These effects, in which our knowledge is poor to say the least, might 
forbid processes even when these processes are kinematically allowed
due to the fluctuations in the energy and momentum. 

\section{The effect of Space-Time fluctuations on the propagation
of high energy particles.}

In this section we summarize the formalism introduced in \cite{noi2} following
\cite{ng1}. The basic points can be listed as follows:
\begin{itemize}
\item{the values of energy (momentum), fluctuate around their average values
(assumed to be the result that in principle could be recovered with 
an infinite number of measurements of the same observable): 
\begin{equation}
 E \approx  {\bar E} + \alpha {{\bar E}^2\over M_P} 
\label{eq:Ebar}
\end{equation}
\begin{equation}
 p \approx  {\bar p} + \beta {{\bar p}^2\over M_P} 
\label{eq:pbar}
\end{equation}
with $\alpha$ and $\beta$ normally distributed variables 
(with $O(1)$ variance) and $p$ the modulus 
of the 3-momentum (for simplicity we assume rotationally invariant 
fluctuations);}
\item{the dispersion relation fluctuates as follows: 
\begin{equation}\label{eq:dis}
P_\mu g^{\mu\nu} P_\nu = E^2-p^2 + \gamma \frac{p^3}{M_P}=m^2
\label{eq:PmuPnu}
\end{equation}
and $\gamma$ is again a normally distributed variable. Eqs. 
(\ref{eq:Ebar}-\ref{eq:PmuPnu}) are assumed to hold for $E,p << M_P$.}  
\end{itemize}
In principle, a theory of Quantum Gravity (QG) should be able to predict
the specific properties of space-time and the dispersion relations written
above, but this approach is clearly out of reach as of today. From this the
need of assuming a gaussian form for the fluctuations introduced above.
It is however worth stressing once more (see also \cite{noi2}) 
that the results that
one gets are not very sensitive to this assumption, namely any symmetric 
distribution with variance $\approx 1$, within a large factor, would
give essentially the same results.
Besides assuming gaussianity, we also {\it assume} that  $\alpha$, $\beta$ 
and $\gamma$ are uncorrelated random variables; again, this assumption 
reflects our ignorance in the dynamics of QG.

The form chosen for the fluctations can be guessed in a variety of
ways, as will be discussed in the last section; notice that we have
chosen a very general form in order to derive our results in a way that 
we consider as unbiased as possible. We also assume  energy momentum
conservation relations on the fluctuating quantities, {\it i.e.}
\begin{equation}
\Sigma \left [ \bar E_i + \alpha_i {\bar E_i^2 \over M_P}\right ]=
\Sigma \left [ \bar E_f + \alpha_f {\bar E_f^2 \over M_P} \right ]
\end{equation} 

\begin{equation}
\Sigma \left [ \bar p_i + \beta_i {\bar p_i^2 \over M_P} \right ]=
\Sigma \left [ \bar p_f + \beta_f {\bar p_f^2 \over M_P} \right ]
\end{equation} 
where the index $i$ ($f$) denotes values of energy and momentum in the initial 
(final) state of a reaction.

\section{Decay of stable particles}

In this section we consider three specific decay channels, that illustrate
well, in our opinion, the consequences of the quantum fluctuations introduced
above. We start with the reaction $$p\to p + \pi^0$$
and we denote with $p$ ($p'$) the momentum of the initial (final) proton, and
with $k$ the momentum of the pion. Clearly this reaction cannot take place 
in the reality as we know it, due to energy conservation. However, since 
fluctuations have the effect of emulating an effective mass of the particles, 
it may happen that for some realizations, the effective mass induced to the 
final proton is smaller than the mass of the proton in the initial 
state, therefore allowing the decay from the kinematical point of view. Since
no conservation law or discrete symmetry is violated in this reaction, it 
may potentially take place. For the sake of clarity, it may be useful to 
invoke as an example the decay of the $\Delta^+$ resonance, which is 
structurally identical to a proton, but may decay to a proton and a pion
according to the reaction $\Delta^+\to p+\pi^0$, since its mass is larger
than that of a proton. From the physical point of view, the effect of the
quantum fluctuations may be imagined as that of {\it exciting} the proton,
inducing a mass slightly larger than its own (average) physical mass.

Following \cite{noi2} we expect to find that for momenta above a given 
threshold, depending on the value of the random variables, the decay may 
become kinematically allowed. In general, the probability for this to 
happen has to be calculated  numerically from the conservation equations 
supplemented by the dispersion relations. 

Although a full calculation is possible, it is probably more instructive
to proceed in a simplified way, in which only the fluctuations in
the dispersion relation of the particle in the initial state are taken
into account. Neglecting the corresponding fluctuations in the final state
should not affect the conclusions in any appreciable way, unless the
fluctuations in the initial and final states are correlated.

In this approximation, the threshold for the process of proton decay to a 
proton and a neutral pion can be written as follows (neglecting corrections 
to order higher than $p/M_P$):

\begin{equation}
\gamma {2 p_{th}^3 \over { M_P}} 
-2 m_{\pi} m_p - m_{\pi}^2 =0,
\end{equation}
with solution
\begin{equation}
p_{th}= \left ({(2m_p m_{\pi} +  m_{\pi}^2 )M_P \over {2 \gamma }} 
\right )^{1\over 3}.
\end{equation}
For negative values of $\gamma$, the above equation has no positive
root; this happens in $50\%$ of the cases. Since the gaussian distribution 
is essentially flat in a small interval around zero, the distribution of 
thresholds for positive $\gamma$ ({\it i.e. } in the remaining 50 $\%$ of the
cases) peaks around the value for $\gamma \approx 1$, meaning that the threshold 
moves almost always down to a value of $\approx 10^{15}$ eV \cite{noi,noi2}; 
essentially the same result holds for generic fluctuations ({\it i.e.} not 
confined to the dispersion relations) affecting only the incident particle,
namely the one with the highest energy.

The reason why the effects of fluctuations are expected to occur at such 
low energies is that in that energy region the fluctuation term
becomes comparable
with the rest mass of the particle. In fact the same concept of rest 
mass of a particle may lose its traditional meaning at sufficiently
high energies \cite{lieu}.

It can be numerically confirmed that {\it independent} fluctuations of  
momenta (and/or of the dispersion relations) of the decay products are more 
likely to make the decay easier rather than more difficult, due to the non 
linear dependence of the threshold on the strength of fluctations:
the probability that the decay does not take place is in fact $\approx 30 \%$. 
In the remaining cases, the decay will occur if the momentum of the initial 
proton is larger than $p_{th}$. The distribution of $p_{th}$ is essentially 
identical to the one reported in \cite{noi2} for other reactions.

All the discussion reported so far remains basically unchanged if similar
reactions are considered. For instance the reaction $p \to \pi^+ n$ is
kinematically identical to the one discussed above. For all these reactions,
we expect that once they become kinematically allowed, the energy loss
of the parent baryon is fast. For the case of nuclei, all the decays that do
not change the nature of the nucleon leave (A,Z) unchanged, so we do
not expect any substantial blocking effect in nuclei.

Another reaction that may be instructive to investigate is the spontaneous 
pair production from a single photon, namely
$$\gamma \to e^+ e^-.$$
In this case, following the calculations described above, we obtain the 
following expression for the threshold:
\begin{equation}
p'_{th}= \left ({4 m_e^2 M_P \over {2 \gamma' }}, 
\right )^{1\over 3}
\end{equation}
and $p'_{th}$ is of the order of $10^{13}$ eV.
Again, if the reaction becomes kinematically allowed, there does not seem 
to be any reason why the reaction should not take place with a rate 
dictated by the typical cross section of electromagnetic interactions. 

Finally, we propose a third reaction that in its simplicity may represent
the clearest example of reactions that should occur in a world in which 
quantum fluctuations behave in the way described above. Let us consider
a proton that moves in the vacuum with constant velocity, and let us consider
the elementary reaction of spontaneous photon emission. In the Lorentz 
invariant world the process of photon emission
is known to happen only in the presence of an external field that 
may provide the conditions for energy and momentum conservation. However, 
in the presence of quantum fluctuations, one can think of the gravitational
fluctuating field as such an external field, so that the particle can in fact 
radiate a photon without being in the presence of a nucleus or some other 
external recognizable field. The threshold for this process, calculated
following the usual procedure, is 
\begin{equation}
p_{th}'' \approx \left ({{m^2 M_P \omega} \over \gamma''}
\right )^{1 \over 4},
\end{equation} 
where $\omega$ is the energy of the photon. This threshold approaches zero 
when $\omega \to 0$: for instance, if $\omega=1$ eV, then $p_{th} \approx 300$ 
GeV for protons and $p_{th} \approx 45$ GeV for electrons. In other words there
should be a sizable energy loss of a particle in terms of soft photons. 
This process can be viewed as a sort of bremsstrahlung emission of a charged
particle in the presence of the (fluctuating) vacuum gravitational potential.

Based on the arguments provided in this section, it appears that all particles
that we do know are stable in our world, should instead be unstable at sufficiently 
high energy, due to the quantum fluctuations described above. In the next section 
we will take a closer look at the implications of the existence of these quantum 
fluctuations, and possibly propose some plausible avenues to avoid
these dramatic conclusions.

\section{Discussion and outlook}

If the decays discussed in the previous section could take place, our
universe, at energies above a few PeV or even at much lower energies 
might be unstable, nothing like what we actually see.
The decays $nucleon\to nucleon+\pi$ would start to be kinematically allowed 
at energies that are of typical concern for cosmic ray physics, while 
the spontaneous emission of photons in vacuum might even start playing
a role at much lower energies, testable in laboratory experiments. 
Without detailed calculations of energy loss rates it is difficult to 
assess the experimental consequences of this process. We are carrying out
these calculations, that will be presented in a forthcoming publication
\cite{noi3}.

For the nucleon decay, the situation is slightly simpler if we assume
that the quantum fluctuations affect only the kinematics but not the
dynamics, an assumption also used in \cite{noi2}. In this case one would expect 
the proton to suffer the decay to a proton and a pion on a time scale of the
same order of magnitude of typical decays mediated by strong interactions.
This would basically cause no cosmic ray with energy above $\sim 10^{15}$
eV to be around, something that appears to be in evident contradiction 
with observations 
\footnote{From a phenomenological point of view, consistency 
with experiments would require either that the variance of the fluctuations 
considered above is ridicolously small ($<10^{-24}$) or, allowing more generic
fluctations $\Delta l \propto l_P (l_P/l)^{\alpha}$, that a fairly large value
for $\alpha$ should be adopted \cite{noi2}.}. 

In the following we will try to provide a plausible answer to these three very 
delicate questions:
\begin{enumerate}
\item If the particles were kinematically allowed to decay, and there
were no fundamental symmetries able to prevent the decay, would it 
take place? 

\item Is the form adopted for the quantum fluctuations correct and if so,
how general is it?

\item If in fact the form adopted for the fluctuations is correct, how
general and unavoidable is the consequence that (experimentally)
unobserved decays should take place?
\end{enumerate}

Although the result that particles are kinematically allowed to
decay is fairly general, the (approximate) lack of relativistic
invariance forbids the computation of life-times \footnote{In fact
life-times can be in principle estimated in approaches in which it is
possible to make transformations between frames \cite{lieu,dsr,jap},
despite the lack of LI.}.
Two comments are in order: first, the phase space for the decays described
above, as calculated in the laboratory frame, is non zero and in fact 
it increases with the momentum of the parent particle. The effect of 
fluctuations can be seen as the generation of an effective (mass)$^2 
\propto p^3/M_P$. A similar effect, although in a slighty different 
context, was noted in \cite{colgla}.
Second, we do not expect dynamics to forbid the reactions:
one must keep in mind that we are considering very small effects, at
momenta much smaller than the Planck scale. For instance the gravitational
potential of the vacuum fluctations is expected to move quarks in a
proton to excited levels, not to change its content, nor the properties 
of strong interactions. 

There is a subtler possibility, which must be taken very seriously in our
opinion, since it might invalidate completely the line of thought illustrated
above, namely that the quantum fluctuations of the momenta of the particles
involved in a reaction occur on time scales that are enormously smaller than 
the typical interaction/decay times. This situation might resemble the
so called Quantum Zeno paradox, where continuously checking for the
decay of an unstable particle effectively impedes its decay. 
This possibility is certainly worth a detailed study, that would however
force one to handle the intricacies of matter in a Quantum Gravity regime. 
We regard this possibility as the most serious threat to the validity of 
the arguments in favor of quantum fluctuations discussed in this paper 
and in many others before it. 

Let us turn out attention toward the question about the correctness and
generality of the form adopted for the momentum fluctuations. It is 
generally accepted that the geometry of space-time suffers profound 
modifications at length (time) scales of the order of the Planck
length (time), and that this leads to the emergence of a minimum
measurable length. This may be reflected in a non commutativity of 
space-time and in a generalized form of the uncertainty principle.
The transition from uncertainty in the length or time scales to uncertainty
in momenta of particles is undoubtly more contrived and deserves some
attention. The expressions in Eqs. \ref{eq:Ebar},\ref{eq:pbar} and 
\ref{eq:PmuPnu} have been 
motivated in various ways \cite{ford,ng1,noi2,lieu,camacho} in previous
papers. For instance, the condition $\Delta l \ge l_P$ implies the following
constraint on wavelengths $\Delta \lambda \ge l_P$, otherwise it would be 
possible to design an experimental set-up capable of measuring distances with 
precision higher than $l_P$. Therefore $\Delta p \propto \Delta (\lambda^{-1})
\propto  l_P p^2$. Similar arguments have been proposed, all based to some
extent on the de Broglie relation $p \propto \lambda^{-1}$. 

There is certainly no guarantee that the de Broglie relation continues
to keep its meaning in the extreme conditions we are discussing, in 
particular in models in which the coordinates and coordinate-momentum 
commutators are modified with respect to standard quantum mechanics and 
the representation of momentum in terms of coordinate derivatives generally 
fails. For instance in a specific (although non-relativistic) example 
\cite{kempf} the existence of a minimum length is shown to imply that
\begin{equation}
p = {2 \over { \pi l_P}} \tan \left ( {\pi l_P \over 2 \lambda} 
\right ).
\end{equation}
In other words, the de Broglie relation may be modified in such a way that
a minimum wavelength corresponds to an unbound momentum.
Notice, however, that we are considering here the effects of these
modifications at length scales much larger than the Planck scale, where the
correction is likely to be negligible. In general, if $p \propto
\lambda^{-1}g(l_P/ \lambda)$ then $\Delta p \propto l_P  p^2 +
p ~O(l_P^2 p^2)$. Hence, we do not expect that the result shown in the
previous Section is appreciably modified.

Last but not least we notice that the fluctuations in the dispersion 
relations can be easily derived from fluctuations of the (vacuum) metric in 
the form given in \cite{camacho}:
\begin{equation}
ds^2 = (1+\phi)dt^2-(1+\psi)d{\bf r}^2
\end{equation}
where $ \phi, ~\psi $ are functions of the position in space-time. 

The fluctuations of the dispersion relation, Eq. \ref{eq:PmuPnu}, follow 
if $\phi \ne \psi$ ({\it i.e.} non conformal fluctuations), assuming at 
least approximate validity of the de Broglie relation; if $\phi=\psi$ a 
much milder modification (O($p m^2/M_P$)) follows.

Having given plausibility arguments in favor of the form adopted for the
fluctuations, at least for the case of non conformal fluctuations, we are 
left with the goal of proving an answer to the last question listed above, 
namely does a decay actually occur once it is kinematically allowed?
Certainly the answer is positive if one continues to assume momentum and
energy conservation, and modifications of these conservation laws with 
random terms of order $O(p^2/M_P)$ do not change this conclusion.
The question then is whether we are justified in assuming energy and momentum
conservation in the form used above. For instance, in the so-called Doubly
Special Relativity (DSR, \cite{dsr}) theories and in general in models with
deformed Poincare' invariance, the conservation relations may be
modified in a non trivial, non additive and non abelian way. For instance, 
in the case of proton decay considered above, momentum conservation
may read as \cite{dsr,jap}
\begin{equation}
{\bf p}_p \approx {\bf p}'_{p}+(1+l_P E'_p) {\bf p}_{\pi}\quad\quad
{\rm or}
\quad\quad 
{\bf p}_p \approx {\bf p}'_{\pi}+(1+l_P E'_{\pi}) {\bf p}_p.
\end{equation}
This certainly makes the probability of being above threshold smaller.
However in order to qualitatively modify our results this probability
should be in fact vanishingly small. For the case of {\it low} energy 
cosmic rays, this probability should be of the order of a typical decay
time divided by the residence time of cosmic rays (mostly galactic at
these energies) in our Galaxy. 

We are led to conclude that allowing for modifications of the conservation 
relations does not appear to improve the situation to the point that  
the strong conclusions derived in the previous section can be avoided.
In the same perspective, cancellation between fixed modifications 
of the dispersion relation and fluctuations (of the same order of 
magnitude) does not seem a viable way to proceed.

It is important however to notice that we have considered the above 
fluctuations as independent. In a full theory (exemplified by DSR models, for
instance) one should take into account the correlations (and possible
cancellations \footnote{In fact in DSR theories the processes described here
should be absent, given the frame independence on which these theories are 
built. We thank Giovanni Amelino-Camelia for having pointed out this fact 
to us.}) between them. In our opinion this analysis is mandatory if
we want to have a clearer idea of the extent of the implications of these
theories. At the level these calculations can be carried 
out at present, we think that there are solid arguments that suggest that 
the fluctuations in the form assumed above imply observational consequences 
which seem to be in serious contradiction with reality.

{\bf Note added in proof}: After submission of the present paper,
analogous considerations appeared in \cite{legal}, where essentially
identical conclusions are reached.

{\bf Acknowledgements}: We warmly thank Piera Luisa Ghia for collaboration 
in an earlier phase and encouragement. We are grateful to Giuseppe Di Carlo 
and Jerzy Kowalski-Glikman for very useful and stimulating discussions.

\end{document}